\documentstyle[11pt,newpasp,twoside,epsf]{article}
\def\edcomment#1{\iffalse\marginpar{\raggedright\sl#1\/}\else\relax\fi}
\def\etal{{\it et al. }}
\reversemarginpar

\begin{document}
\title{Observational Properties of Diffuse Halos in Clusters}
 \author{Luigina Feretti}
\affil{Istituto di Radioastronomia CNR, Via Gobetti 101, 40129 Bologna, 
Italy}

\begin{abstract}
 The number of known diffuse radio sources in clusters of
galaxies has grown in recent years, making it possible to derive
statistical properties of these sources and of the hosting clusters.
We show that diffuse sources are associated with X-ray luminous 
clusters which have undergone recent merger processes. 
The radio and X-ray structures are often similar, and  correlations
are found between radio and X-ray parameters. This is indication 
of a link between the diffuse relativistic and thermal plasma in
clusters of galaxies. 
\end{abstract}

\section{Introduction}

It is well established that an important component
of the intergalactic medium (IGM) in clusters 
and groups of galaxies  is the hot gas, 
observed in X-rays and characterized by 
temperatures in the range $\sim$5-10 keV, by a central 
density of $\sim$10$^{-3}$ cm$^{-3}$ and by a
density distribution approximated by a beta model (Cavaliere \& Fusco-Femiano
1981).
In addition, magnetic fields are wide spread in clusters (e.g. Eilek 1999),
as deduced by Rotation Measure arguments, and relativistic electrons
may be common (Sarazin \& Lieu 1998).
These two non-thermal components can be directly revealed 
in some clusters by the presence of diffuse extended 
radio sources, which are related to the intergalactic medium,
rather than to a particular cluster galaxy.
However, diffuse sources seem not to be a general property
of the IGM.

The importance of these sources is that they represent large scale 
features, which are related to other cluster
properties in the optical and X-ray domain, and are
thus directly connected to the cluster history
and evolution. 
 In this paper, the observational properties  of diffuse cluster sources 
and of their host clusters are   outlined.
Intrinsic parameters are calculated with H$_0$ = 50 km s$^{-1}$ Mpc$^{-1}$
and q$_0$ = 0.5.

\section{Classification and radio properties}

The diffuse source Coma C in the Coma cluster (Fig. 1, left panel), 
discovered  30 years ago 
(Willson 1970), is the prototypical example of a cluster {\it radio halo}.
The radio halo is located at  the cluster center, it
has a steep radio spectrum ($\alpha \sim
1.3$) and is extended $\sim$ 1 Mpc (Giovannini \etal 1993). 
Another example of cluster-wide halo, 
associated with the cluster A~2255,
is shown in the right panel of Fig. 1.
An additional diffuse source, 1253+275, is detected at the Coma
cluster periphery, which might be connected to
the cluster halo by a very low-brightness radio bridge
(Giovannini \etal 1991). This source and similar diffuse sources 
located in peripheral cluster regions are referred to  as
{\it radio relics} in the literature.
This name may be misleading, since it can  also be used to indicate 
dying radio galaxies, without active nucleus, as
B2~1610+29 in A~2162 (Parma et al. 1986). I would suggest 
{\it peripheral halos}, whereas Ron Ekers suggested {\it radio flotsam}.
Since the interpretation of these sources is still unclear, 
I will use here the name {\it relics}, for homogeneity with the literature.

\begin{figure}
\plottwo{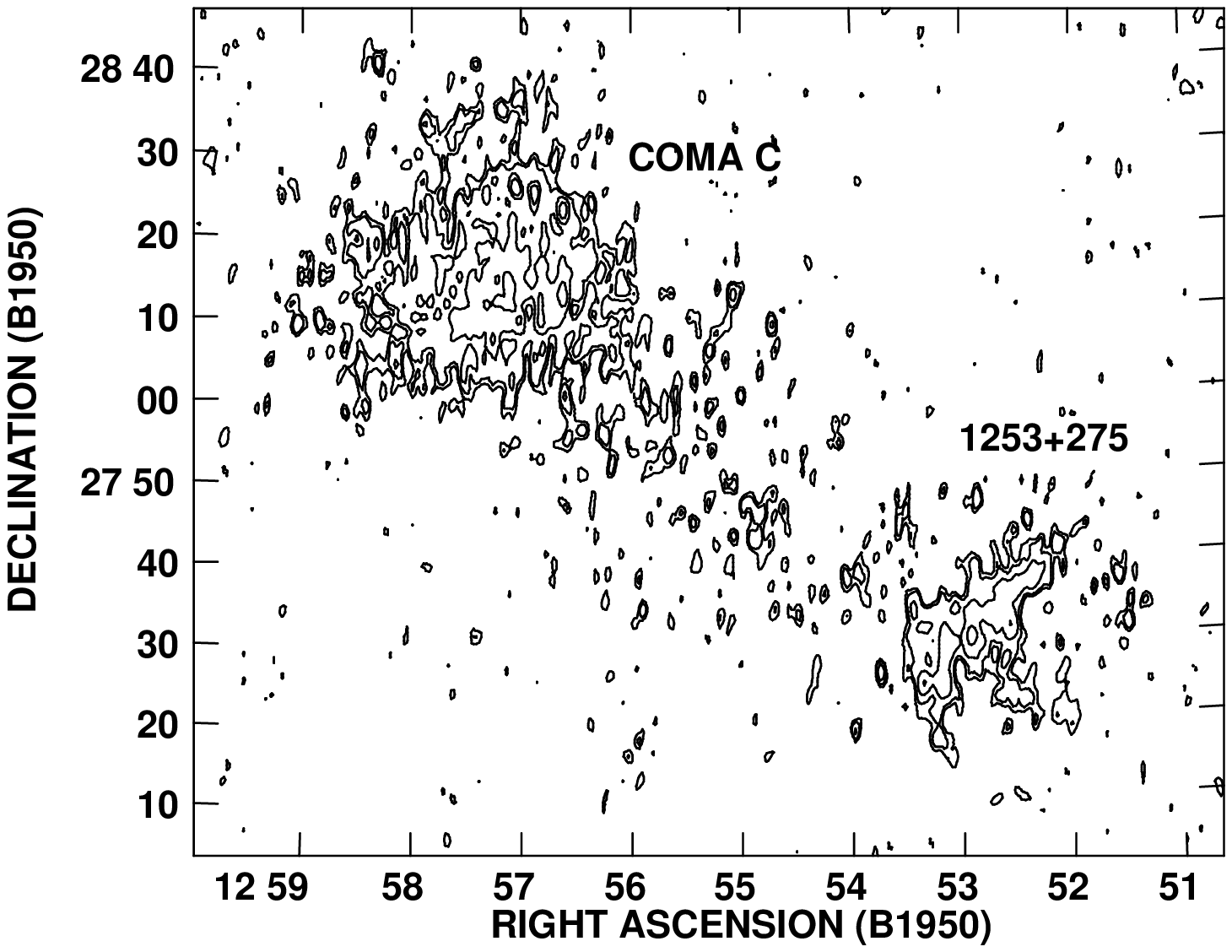}{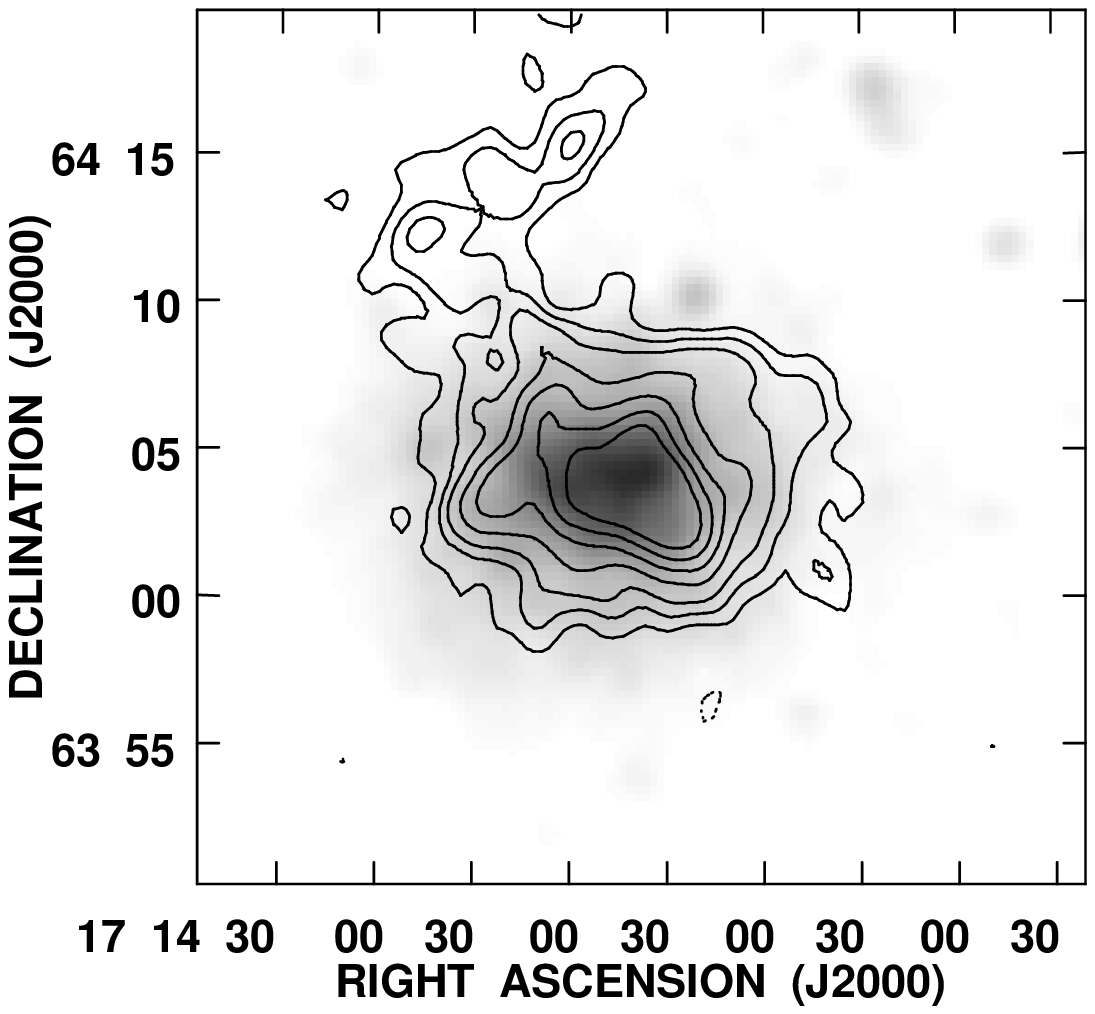}
\caption
{Examples of diffuse sources in clusters, after subtraction of the
discrete sources. {\it Left panel:} Radio map of Coma at 90 cm, 
with angular resolution of 55$^{\prime\prime}$ $\times$ 
125$^{\prime\prime}$ (HPBW, RA $\times$ DEC). Contour levels 
are 2.5, 4, 8 16  mJy/beam. The two diffuse sources are easily 
visible. {\it Right panel:} Radio image of the diffuse emission 
in A~2255 at 90 cm (contours)
superimposed on the greyscale X-ray ROSAT image. Contour levels are
-4, 4, 7 , 10, 15, 20 25, 30 mJy/beam, with HPBW=89$^{\prime\prime}$
$\times$84$^{\prime\prime}$(@6$^{\circ})$.}
\end{figure}

Radio halos and relics show  
low surface brightness ($\sim \mu$Jy/arcsec$^2$ at 20 cm) 
and steep radio spectrum.  
Their  detection is limited by the surface brightness
sensitivity coupled with the high resolution needed
to  separate  such sources
from the  embedded discrete sources.
Because of their steep spectrum, they are better 
detected at lower frequencies.

\begin{table}
\caption{Clusters with well studied halos and relics}
\begin{tabular}{ccccccccc}
\hline
 Cluster & z & P$_{1.4}$  & L.S.  & L$_{X bol}$ & T    &  Dist & Class  &  \\
         &   & 10$^{23}$ W Hz$^{-1}$  & Mpc   & 10$^{44}$ erg s$^{-1}$    & keV  & Mpc     \\ 
\hline
A85    &   0.0555 &  6.26  & 0.48  &  19.52          & 5.1  & 0.54      & R &  \\
A115   &   0.1971 &  255.5 & 1.88  &  31.09          & 4.9  & 0.93      & R &  \\
A520   &   0.2030 &  62.1  & 1.08  &  37.35          & 8.5 & -         & H &  \\    
A610   &   0.0956 &  7.65  & 0.57  & -               & -    & 0.71      & R &  \\ 
A665   &   0.1818 &  66.5  & 2.13  &  41.72          & 8.5 & -         & H &  \\   
A773   &   0.2170 &  22.3  & 0.83  &  35.10          & 8.6 & -         & H &   \\   
A1300  &   0.3071 &  92.4  & 0.58  &  47.63          & 10.5 & -         & H &   \\   
       &          &  92.4  & 0.95  &  47.63          & 10.5 & 0.80      & R &   \\
A1367  &   0.0216 &  0.71  & 0.29  &   2.87          & 3.5  & 0.83      & R &   \\
A1656  &   0.0232 &  15.0  & 1.09  & 20.42          &  8.2 & -         & H &   \\
       &          &  7.03  & 1.17  & 20.42           & 8.2  & 2.72      & R &  \\ 
A2163  &   0.2080 &  306.2 & 2.60  & 132.91          & 14.2 & -         & H &  \\
A2218  &   0.1710 &  12.2  & 0.52  & 21.96           &  6.9 & -         & H &  \\
A2255  &   0.0809 &  12.6  & 1.23  & 12.42           & 5.4  & -         & H &  \\
       &          &  3.51  & 0.98  & 12.42           & 5.4  & 1.23      & R &  \\
A2256  &   0.0581 &  4.48  & 1.11  & 18.39           & 7.5 & 0.59      & R &  \\
       &          &  37.3  & 1.19  & 18.39           & 7.5 & -         & H &   \\
A2319  &   0.0555 &  20.8  & 1.41  & 39.74           & 9.7 & -         & H &   \\
A2744  &   0.3080 &  241.6 & 1.81  & 62.44           & 11.0 & -         & H &   \\
       &          &  74.3  & 1.84  & 62.44           & 11.0 & 1.91      & R &   \\
A3667  &   0.0552 &  323.1 & 2.63  & 22.70           & 7    & 2.45      & R &   \\
CL0016+16 &   0.5545 &  89.6  & 1.11  & 28.13           & 8.2 & -      & H &   \\
1E0658-56 &   0.296  &  312.0 & 1.96  & 140             & 14.5 & -      & H &   \\
\hline
\end{tabular}
Caption. Col. 1: cluster name; Col. 2: cluster redshift; Col. 3: radio power
of the diffuse source at 1.4 GHz; Col. 4: largest linear size of the diffuse source;
Col. 5: cluster X-ray bolometric luminosity; Col 6: cluster temperature
obtained by averaging values in the literature; 
Col. 7: projected distance of the diffuse source from the cluster center;
Col. 8: source classification, H=halo, R=relic. 
\end{table}

\begin{figure}
\plotone{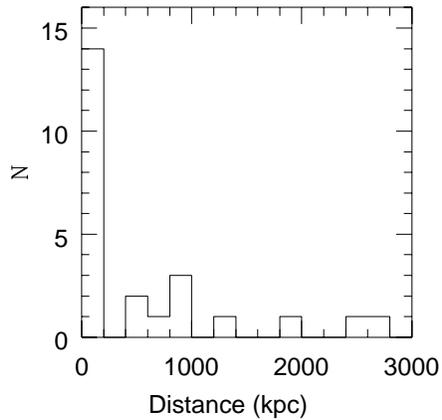}
\caption{Distribution of projected
distances of the diffuse sources from the cluster center.}
\end{figure}

Until recently,  the number of halos and relics was small, thus
these sources were considered to be rare.
This is no longer true. 
Thanks to the better sensitivity of radio telescopes
and to the existence of deep surveys, 
more than 30 clusters hosting diffuse sources are known today. 
For 18 of them (see Table 1) the presence of diffuse radio
emission is well established and good radio data are available
either from the literature or from new observations
(Govoni \etal in preparation).
It is remarkable the existence in some clusters of more than one 
diffuse source.

The  sizes of halos are  typically larger than 1 Mpc. 
Peripheral relics are elongated in shape, and the distribution of their
largest sizes is not  statistically different from that of halos
on a Kolmogorov-Smirnov (KS) test. 

The distribution of projected distances from the cluster center (Fig. 2)
demonstrates that  the diffuse sources are not located 
at random positions in the clusters, i.e. central  
halos are likely to be really at the cluster center and not simply
projected onto it.

Radio powers are of the order of 10$^{24}$-10$^{25}$ W Hz$^{-1}$ at 1.4 GHz.
In a radio size-radio power diagram, the diffuse radio sources 
follow the same correlation of the  radio galaxies (e.g.
Ledlow \etal 2000), lying in the upper part of the plot.

Minimum energy densities in diffuse sources are between
$\sim$ 5 $ $10$^{-14}$ and 2 10$^{-13}$ erg cm$^{-3}$.
This implies that the pressure of relativistic electrons is much lower
than that of the thermal plasma.
Equipartition magnetic fields are about $\sim$0.1-1$\mu$G.
These values can be compared with independent estimates
from Rotation Measure arguments and from Inverse Compton
X-ray emission, to determine if these radio sources are at the
equipartition. 

\section{Occurrence}

To derive the frequency of radio halos and relics we need systematic radio
information on complete samples of clusters.
Giovannini \etal (1999) used the NRAO Vla Sky Survey (NVSS) to search
for diffuse sources associated with  clusters of galaxies from the
sample of Ebeling \etal (1996). This sample is 
complete down to an X-ray  flux of 5 10$^{-12}$
erg cm$^{-2}$ s$^{-1}$ in (0.1-2.4) keV, for redshifts $<$0.2
and for galactic latitude $\mid$b$\mid>$20$^{\circ}$. 
Moreover, it contains some clusters at higher
redshift and at lower galactic latitude.

The total detection rate of clusters with halos + relics in the 
complete sample is
of 5\% to 10\%, and the relative occurence of halos and relics is
 similar  (taking into account the
uncertain detections).  The detection rate
increases with the X-ray luminosity (Giovannini \etal 2000), 
being of $\sim$40\%  in clusters with X-ray
luminosity  larger than 10$^{45}$ erg s$^{-1}$.
In particular, the detection rate of  central radio halos is
more strongly dependent on the X-ray luminosity.

The clusters hosting a diffuse source have a
significantly higher X-ray luminosity than
clusters without a diffuse source ($>$ 99.9\%
confidence level with a KS test).
The detection rate in the high redshift
sample is fully consistent with the previous results.
These results are consistent with those of Owen \etal (1999).

\section{Properties of the host clusters}

An important property derived in the previous section 
is that the clusters hosting diffuse sources are more 
X-ray luminous and consequently they have a high temperature and
a large mass. Values larger than 1.5 10$^{14}$ M$_{\odot}$ are found
 for the total (gravitating) mass within 0.5 Mpc, 
with gas fraction ranging from 10\% to 20\%.
This is consistent with the serendipitous detection
of halos observed during the attempts
to detect  Sunyaev-Zeldovich effect in massive high redshift clusters. 

In previous studies (e.g. Feretti 1999),  radio halos have 
been found to be often associated with clusters showing
indication of merger processes from X-ray and optical structure, 
and from X-ray temperature gradients.
This is confirmed by  statistical studies, according
to the following arguments: 
\par\noindent
- X-ray images of clusters with halos and relics show
the presence of substructures and distortions in the brightness
distribution (Schuecker \& B\"ohringer 1999), which can be interpreted as
the result of subclumps interaction;
\par\noindent
- clusters with halos do not have a strong cooling flow
(e.g. Feretti 1999). This is further indication that a cluster has undergone 
a recent merger, as cooling flow and irregular cluster structure
tend to be anticorrelated (Buote \& Tsai 1996) and
a strong merger process is expected to disrupt a cooling flow
(Peres \etal 1998); 

\begin{figure}
\plottwo{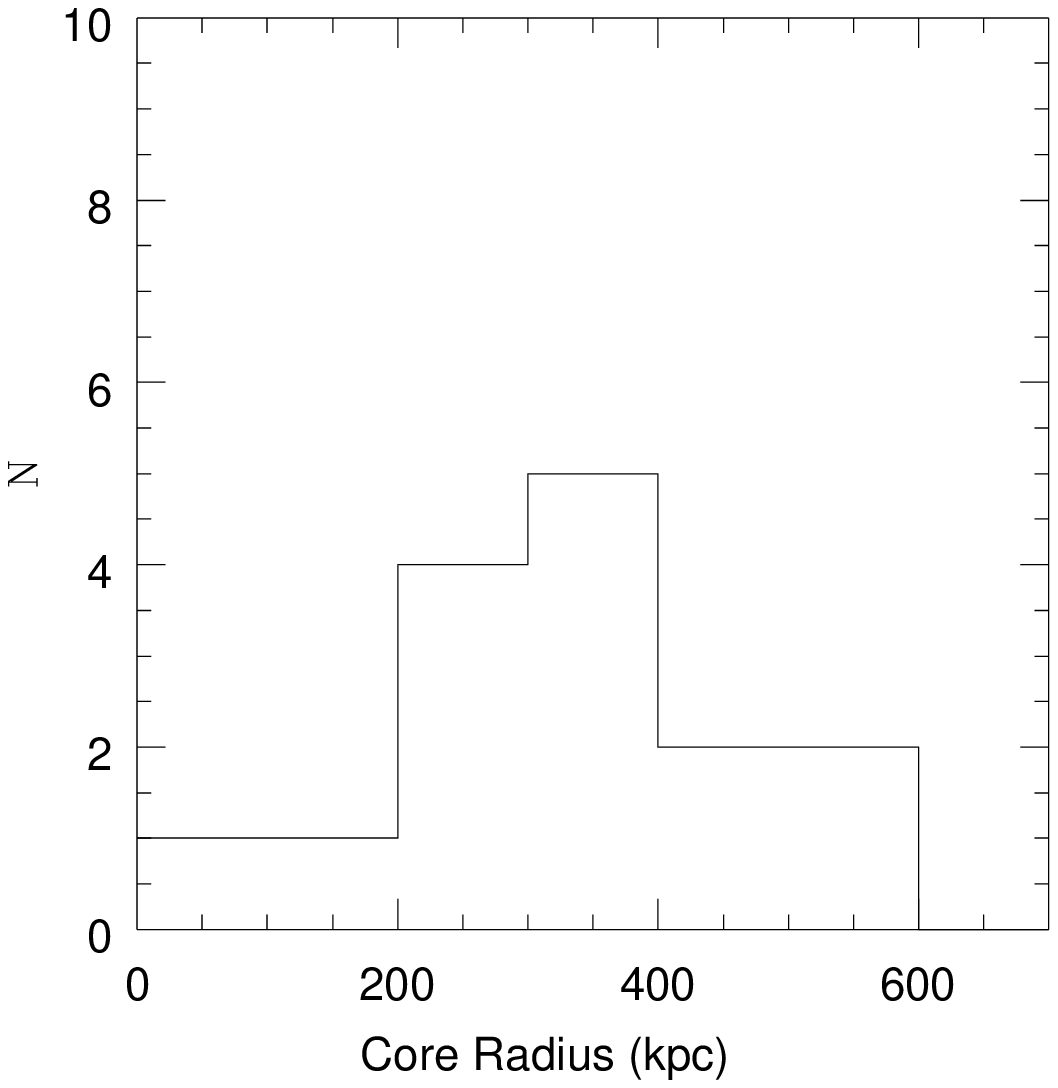}{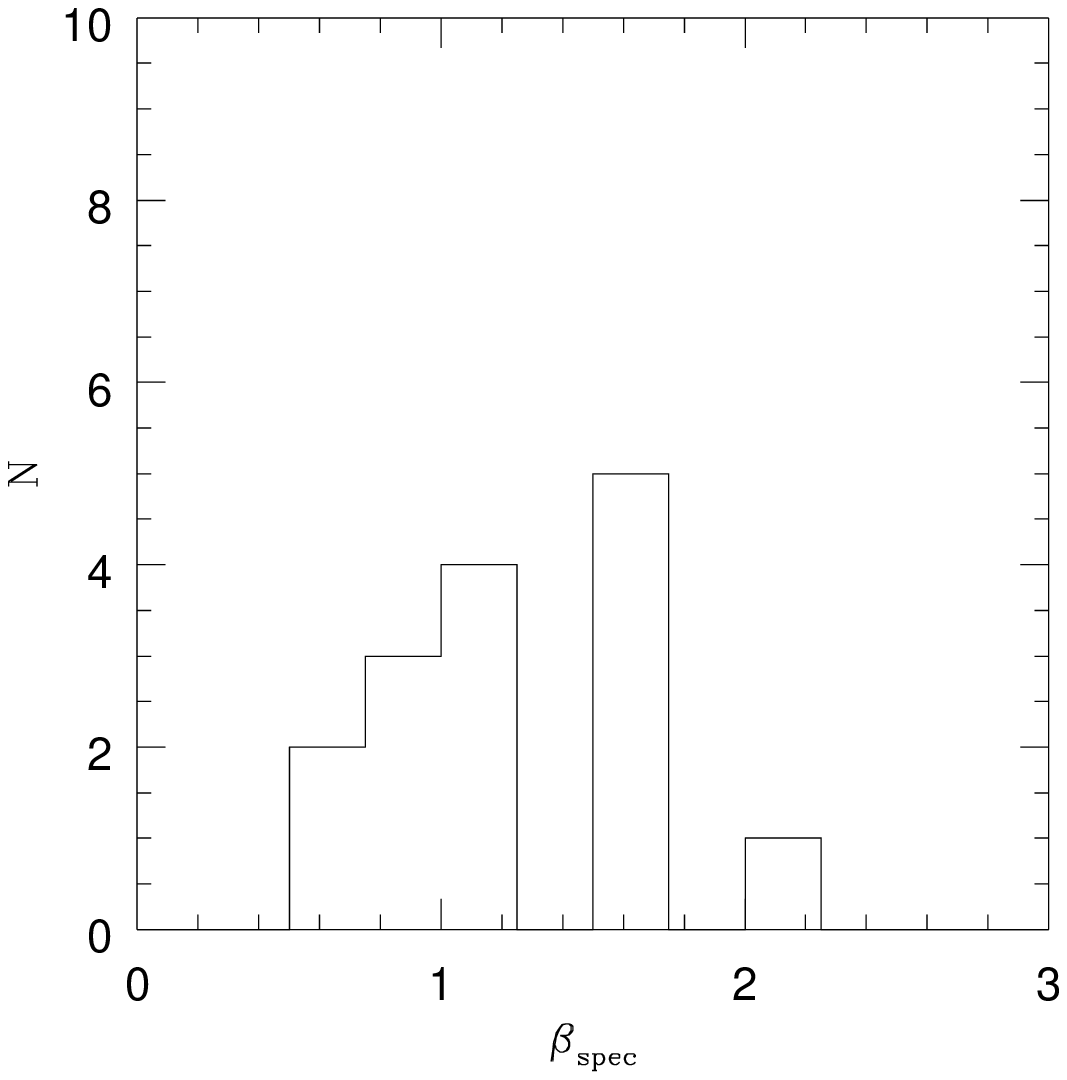}
\caption{{\it Left panel:} Distribution of X-ray core radii of clusters
with halos and relics. {\it Right panel:} Distribution of
spectroscopic $\beta$ of clusters with halos and relics.
}
\end{figure}

\par\noindent
- the X-ray  core radii of clusters with halos/relics in Table 1 
(see Fig. 3, left panel) are significantly larger 
($>$99\% level using a KS test)
than those of clusters classified as single/primary
by Jones anf Forman (1999). According to the last authors,
the large core radius
clusters are  multiple systems in the process of merging and
hotter clusters tend to have larger core radii;
\par\noindent
- for the clusters of Table 1 with optical and X-ray information,
the values of spectroscopic $\beta$ are on average larger 
than 1 (Fig 3, right panel), 
indicating the presence of substructure (Edge \& Stewart, 1991);
\par\noindent
- clusters with halos and relics have larger 
distances to their next neighbours compared to ordinary
clusters with similar X-ray
luminosity, i.e. similar cluster mass (Schuecker and B\"ohringer 1999). 
The fact that they appear more isolated gives additional
support to the idea that recent merger events in halo 
clusters lead to a depletion of the nearest neighbours.

\begin{figure}
\plottwo{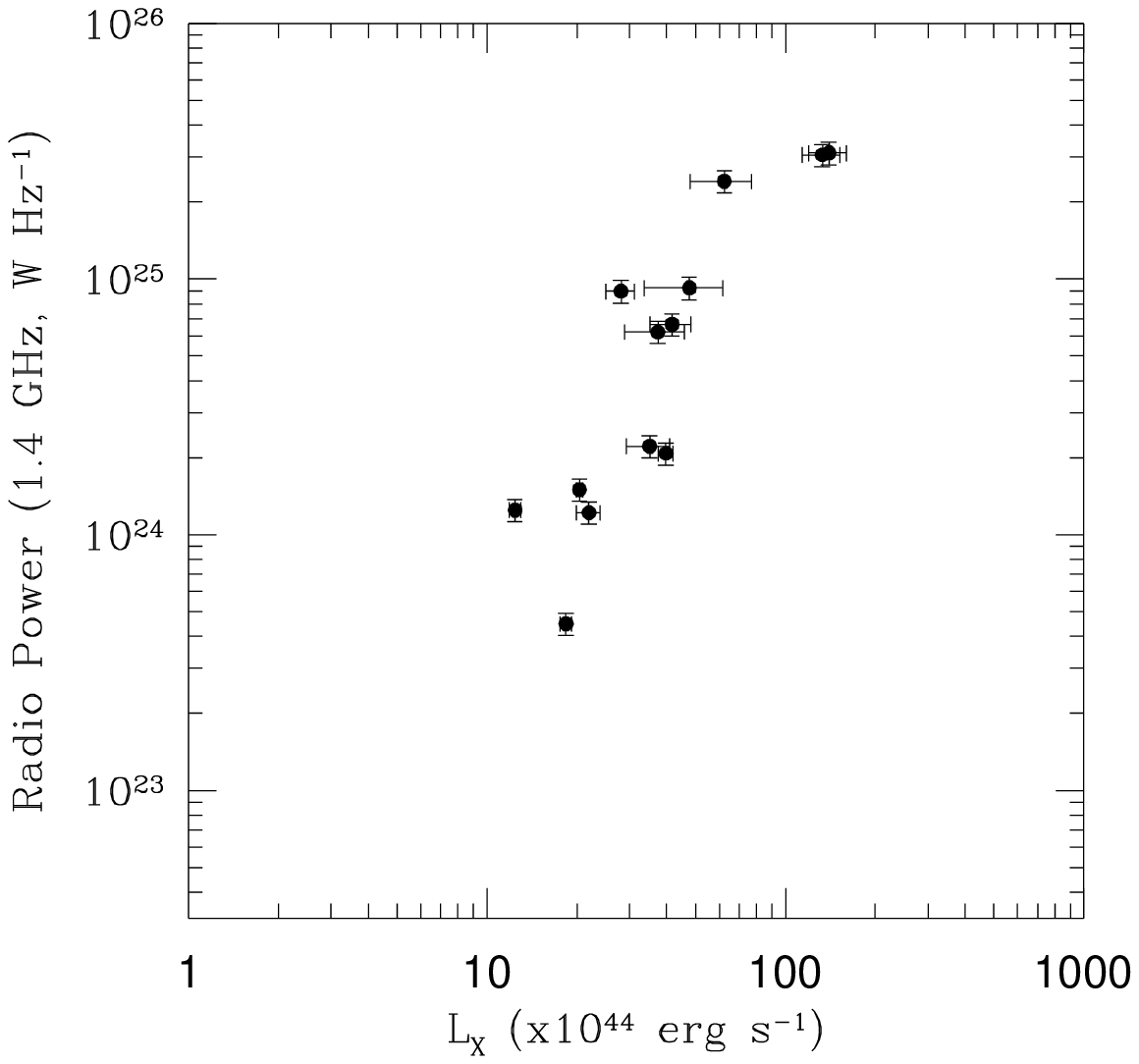}{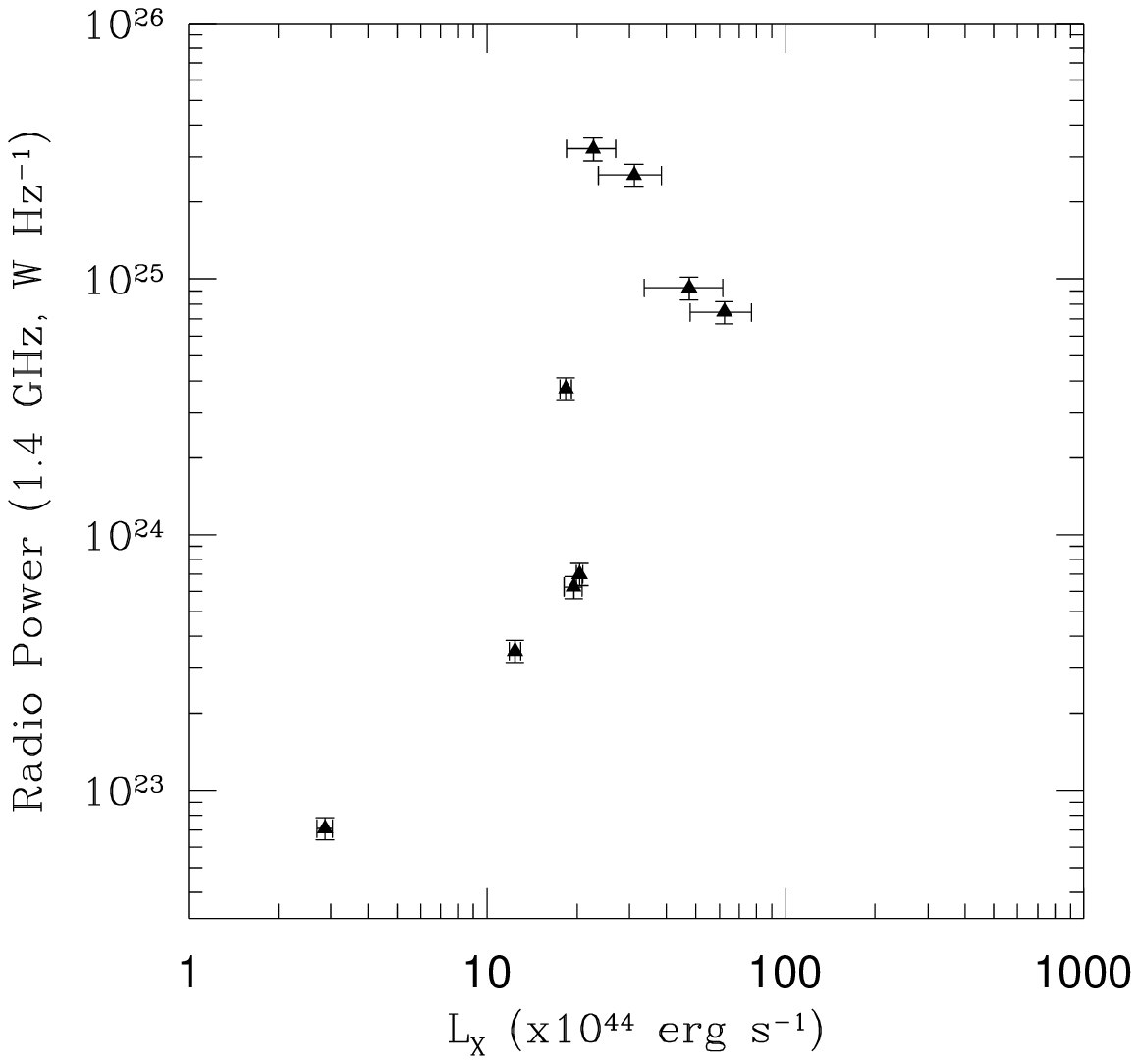}
\caption{Monochromatic radio power versus 
cluster bolometric X-ray luminosity 
for halos (left panel) and relics (right panel).}
\end{figure}

\begin{figure}
\plottwo{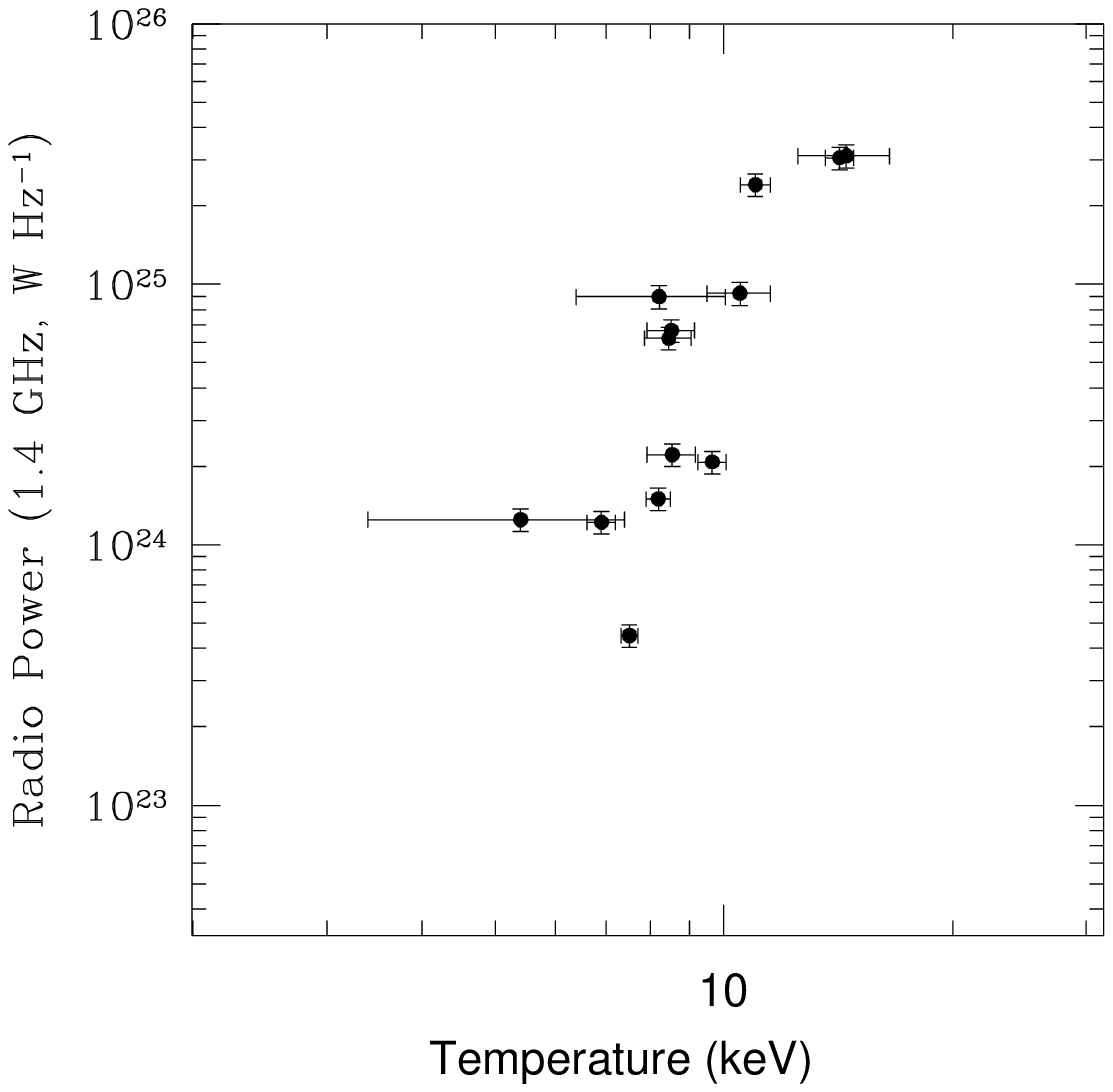}{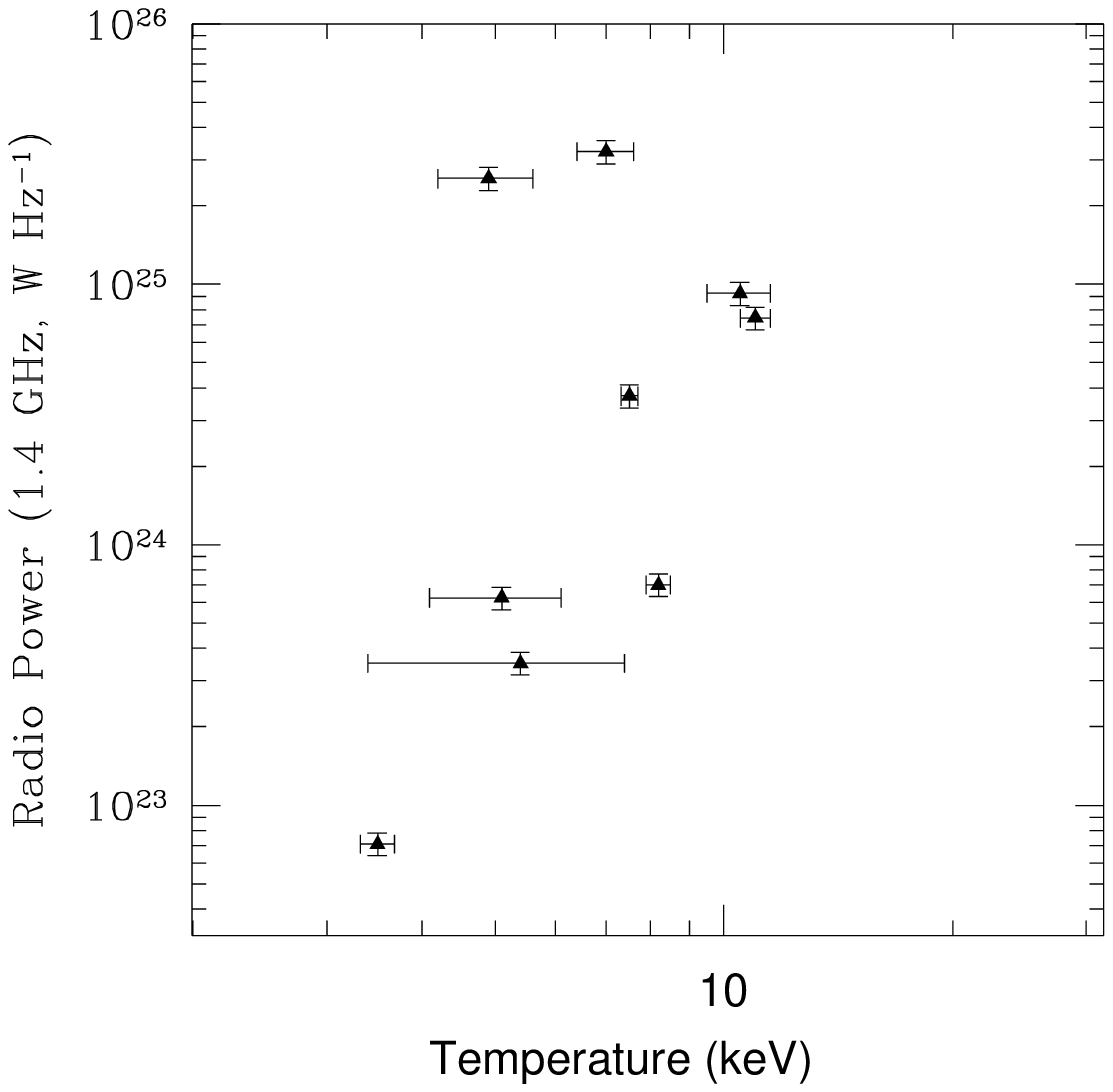}
\caption{Monochromatic radio power versus cluster temperature 
for halos (left panel) and relics (right panel).}
\end{figure}

\begin{figure}
\plottwo{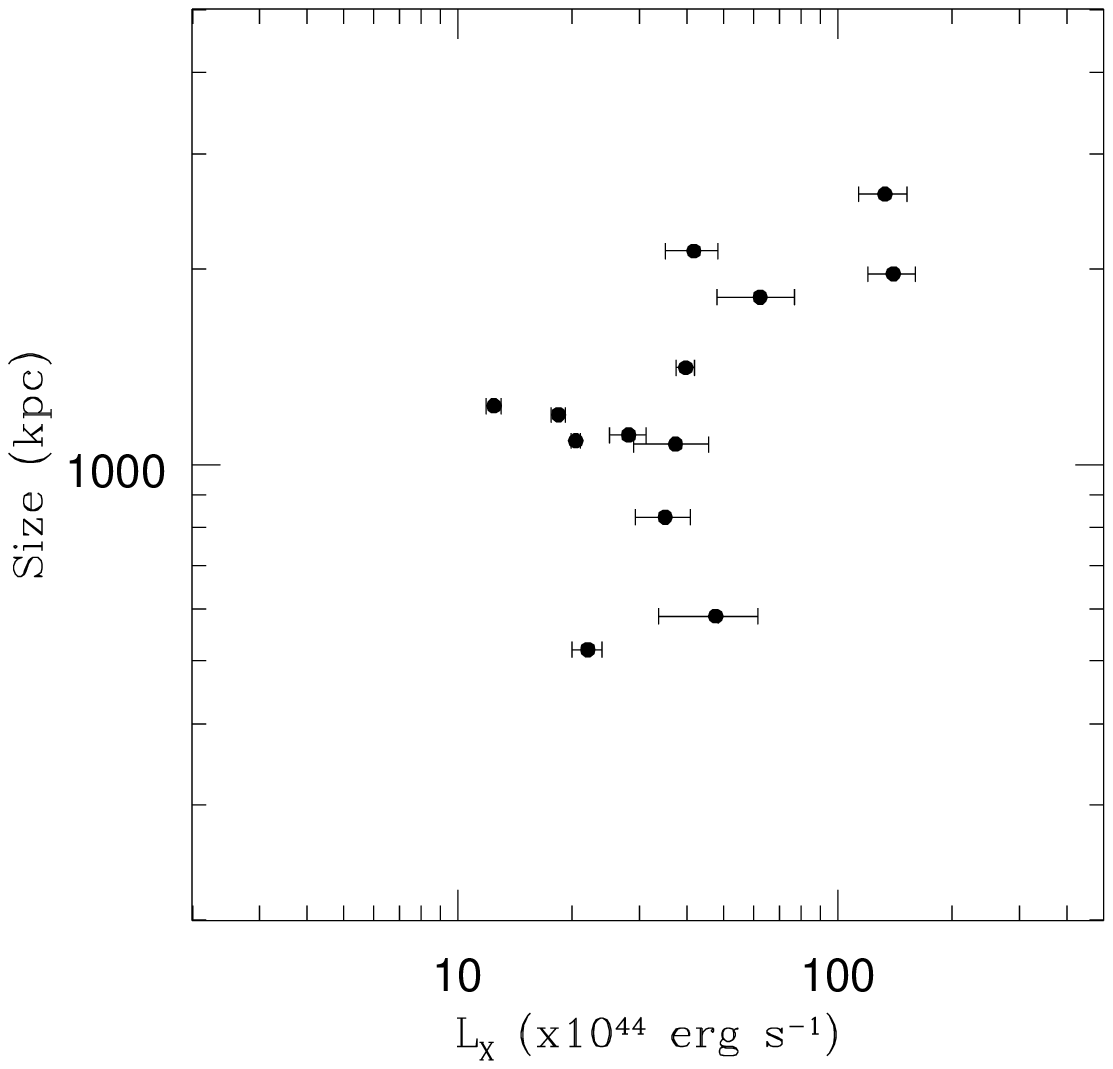}{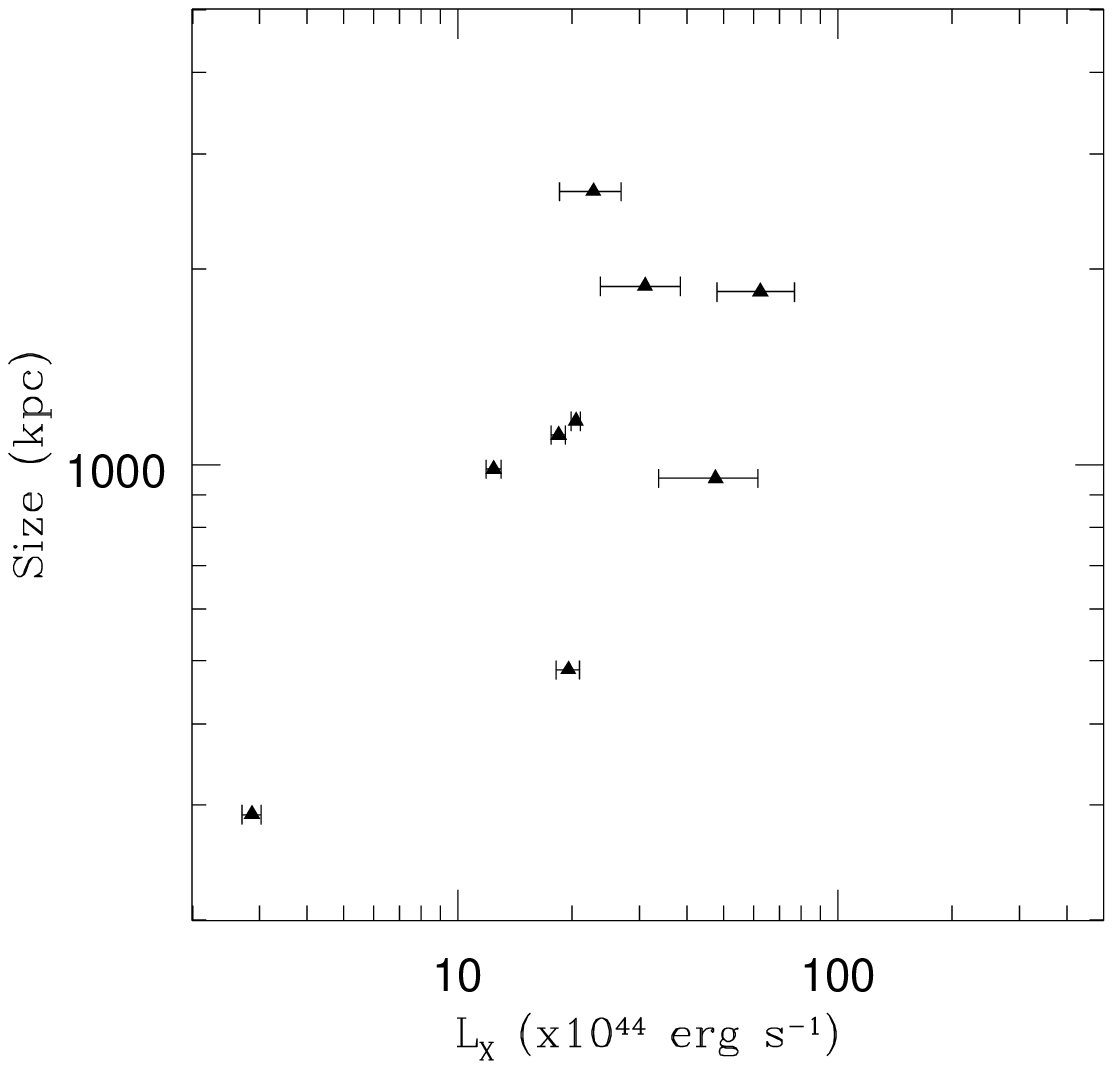}
\caption{Radio largest linear size versus cluster 
bolometric X-ray luminosity
for halos (left panel) and relics (right panel).}
\end{figure}

In conclusion, there seems to be convincing evidence that 
diffuse sources are preferentially associated with high X-ray luminosity
clusters with mergers. To our
knowledge, however, not all the X-ray luminous merging clusters
host a diffuse source, but this point needs further investigation.

\section {Link between relativistic and thermal plasma}

Radio structures of halos show in many cases close similarity to 
the X-ray structures, suggesting
a causal connection between the hot and relativistic plasma.
An example is given in Fig. 1 (right panel), which shows that 
the central radio
halo in A~2255 is elongated in the E-W direction as the X-ray gas.
Other examples are the radio halos
in A~665, A~1300, A~2218, A~2319 (see Feretti 1999 for 
comments on individual clusters) and in 1E0658-56 (Liang 1999).

The correlation visible  between the monochromatic radio power 
at 1.4 GHz and the bolometric 
X-ray luminosity for the  clusters of Table 1 (Fig. 4) implies
a  correlation between radio power
and cluster temperature (Fig. 5),  also derived by Liang (1999).
Since the cluster X-ray luminosity and mass are correlated as well as
the temperature and mass, it follows that the
halo radio power also correlates with mass. The correlations
must be taken into account in model of radio halo formation. It
is of particular interest to understand whether the formation of 
a diffuse source is affected by the high cluster temperature or by 
the large cluster mass. 

A correlation seems to exist also between the largest
radio size of diffuse sources and the cluster X-ray luminosity
(Fig. 6), and goes in the direction of  more X-ray luminous clusters  
hosting larger diffuse sources.
All the correlations are  more evident for radio halos 
than for relics.

Another  link between radio and X-ray is represented by the 
detection of hard X-ray emission produced by the IC scattering 
of relativistic electrons with the microwave background photons.
The hard X-ray emission  detected  in the Coma
cluster (Fusco-Femiano \etal 1999) and in A~2256 (Fusco-Femiano \etal 2000)
has been interpreted in this way.

\section {Discussion and Conclusions}

It is evident from the observational results that
central halos are strictly related to the presence
of recent mergers.
The cluster merger can provide energy for the electron reacceleration
and magnetic field amplification. However, also
the high X-ray luminosity is relevant for the formation of
halos. This seems to imply
that the dynamical history of the cluster, i.e. 
the formation process of a massive hot cluster 
is crucial to trigger  a halo. 
This scenario would explain why not all merging clusters
host a halo (at least 50\% of clusters show mergers, Jones \& Forman
1999, while less than 5\% of clusters show central halos).

Brunetti \etal (2000) suggested a two-phase model, 
successfully applied to Coma C, which includes: 
i) a phase of injection of relativistic electrons 
in the cluster volume from starburst galaxies,
AGNs, and strong shocks during the cluster formation; 
ii)  a phase of particle reacceleration by recent mergers. 
The Brunetti \etal model can be generalized to the other halos, and it
would account for the presence of halos in X-ray luminous
clusters, since the hot massive clusters should have had 
a more efficient injection phase.
 
It still debated if radio halos
and relics have a common origin and evolution, or if they
should be considered as different classes of sources.
Several clusters host both a central and a peripheral halo, thus
favouring a common origin and nature.
Peripheral relics, like halos,  are associated with merging clusters. 
However, the clusters are less luminous in X-ray and the radio-X-ray
correlations are weaker. A current hypothesis is that 
radio relics have a different nature, and
may be reaccelerated by shock waves of an ongoing
merger event, or from shocks during the structure
formation of the Universe (Ensslin 2000).

\acknowledgments

I would like to thank the scientific and local organizing committee for
organizing such an enjoyable and useful conference.
I am grateful to Isabella Gioia and Gabriele Giovannini for helpful
discussions.

\end{document}